\documentclass{ws-ijmpa}
\usepackage[super,compress]{cite}
\usepackage{graphicx}
\usepackage{wasysym}
\begin{document}
%\markboth{Authors' Names}
%{Instructions for Typing Manuscripts (Paper's Title)}

%%%%%%%%%%%%%%%%%%%%% Publisher's Area please ignore %%%%%%%%%%%%%%%
%
\catchline{}{}{}{}{}
%
%%%%%%%%%%%%%%%%%%%%%%%%%%%%%%%%%%%%%%%%%%%%%%%%%%%%%%%%%%%%%%%%%%%%

\title{FUTURE DIRECTIONS IN THE MICROWAVE CAVITY SEARCH FOR DARK MATTER AXIONS
%\footnote{For the title, try not to use more than 3 lines. Typeset the title in 10 pt roman, uppercase and
%boldface.}
}

\author{ T.M. SHOKAIR, J. ROOT, K.A. VAN BIBBER}
\address{Department of Nuclear Engineering, University of California Berkeley, 4153 Etcheverry Hall\\
Berkeley, California 94720, United States\\
%Group, Laboratory, Address\\
%City, State ZIP/Zone, Country\\
%second\_author@domain\_name
shokair@berkeley.edu
}

\author{B. BRUBAKER, Y.V. GUREVICH, S.B. CAHN, S.K. LAMOREAUX}
\address{Department of Physics, Yale University, 217 Prospect Street\\ 
New Haven, Connecticut 06511, United States
%Group, Laboratory, Address\\
%City, State ZIP/Zone, Country\\
%second\_author@domain\_name
}

\author{M.A. ANIL, K.W. LEHNERT, B.K. MITCHELL, A. REED}
\address{Department of Physics, Colorado University, 390 UCB\\ 
Boulder, Colorado 80309, United States
%Group, Laboratory, Address\\
%City, State ZIP/Zone, Country\\
%second\_author@domain\_name
}

\author{G. CAROSI
%\footnote{
%Typeset names in 8 pt roman, uppercase. Use the footnote to indicate the
%present or permanent address of the author.}
}
\address{Physics Division, Lawrence Livermore National Laboratory, 7000 East Avenue\\ 
Livermore, California 94551, United States\\
%University Department, University Name, Address\\
%City, State ZIP/Zone, Country
%\footnote{State completely without abbreviations, the
%affiliation and mailing address, including country. Typeset in 8 pt
%italic.}\\
}
\maketitle

\begin{history}
\received{3 Februrary 2014}
%\revised{Day Month Year}
\end{history}

\begin{abstract}
The axion is a light pseudoscalar particle which suppresses CP-violating effects in strong interactions and also happens to be an excellent dark matter candidate.   Axions constituting the dark matter halo of our galaxy may be detected by their resonant conversion to photons in a microwave cavity permeated by a magnetic field.  The current generation of the microwave cavity experiment has demonstrated sensitivity to plausible axion models, and upgrades in progress should achieve the sensitivity required for a definitive search, at least for low mass axions.  However, a comprehensive strategy for scanning the entire mass range, from 1-1000 $\mu$eV, will require significant technological advances to maintain the needed sensitivity at higher frequencies.  Such advances could include sub-quantum-limited amplifiers based on squeezed vacuum states, bolometers, and/or superconducting microwave cavities.  The Axion Dark Matter eXperiment at High Frequencies (ADMX-HF) represents both a pathfinder for first data in the 20-100 $\mu$eV range ($\sim$5-25 GHz), and an innovation test-bed for these concepts.

%The abstract should summarize the context, content
%and conclusions of the paper in less than 200 words. It should
%not contain any references or displayed equations. Typeset the
%abstract in 8 pt roman with baselineskip of 10 pt, making
%an indentation of 1.5 pica on the left and right margins.

\end{abstract}

\ccode{PACS numbers:14.80.Va, 95.35.+d, 84.30.Le, 85.25.Cp, 52.77.Dq,}

%\tableofcontents

\section{Introduction}

The axion, a light hypothetical pseudoscalar, arises as a natural solution to the strong CP problem,\cite{1,2,3} and now appears to be a ubiquitous feature within all string theories.\cite{4}  A sufficiently light axion also represents a compelling dark-matter candidate, with a density  relative to the critical density of the universe given by\cite{5}

\begin{equation}
\Omega_a = \left(\frac{6\mathrm{\ \mu eV}}{m_a}\right)^{\frac{7}{6}}.\\
\label{equ:1}
\end{equation}

%Equations should be referred to in abbreviated form,
%e.g.~``Eq.~(\ref{diseqn})'' or ``(2)''. In multiple-line
%equations, the number should be given on the last line.

\noindent An axion of $m_a \approx 20 \ \mu$eV (within a factor of $\sim$2) would thus account for the entire dark matter density of the universe, $\Omega_m \approx 0.26$.  Axions of mass $<10^{-6}$ would overclose the universe, while axions with mass exceeding a few meV would have quenched the neutrino pulse observed from SN1987a. This leaves $m_a = 10^{-6} - 3\times10^{-3}$ eV as the allowed mass range, subject to a number of theoretical uncertainties.\\  

\begin{figure}[b]
\centerline{\includegraphics[width=12cm]{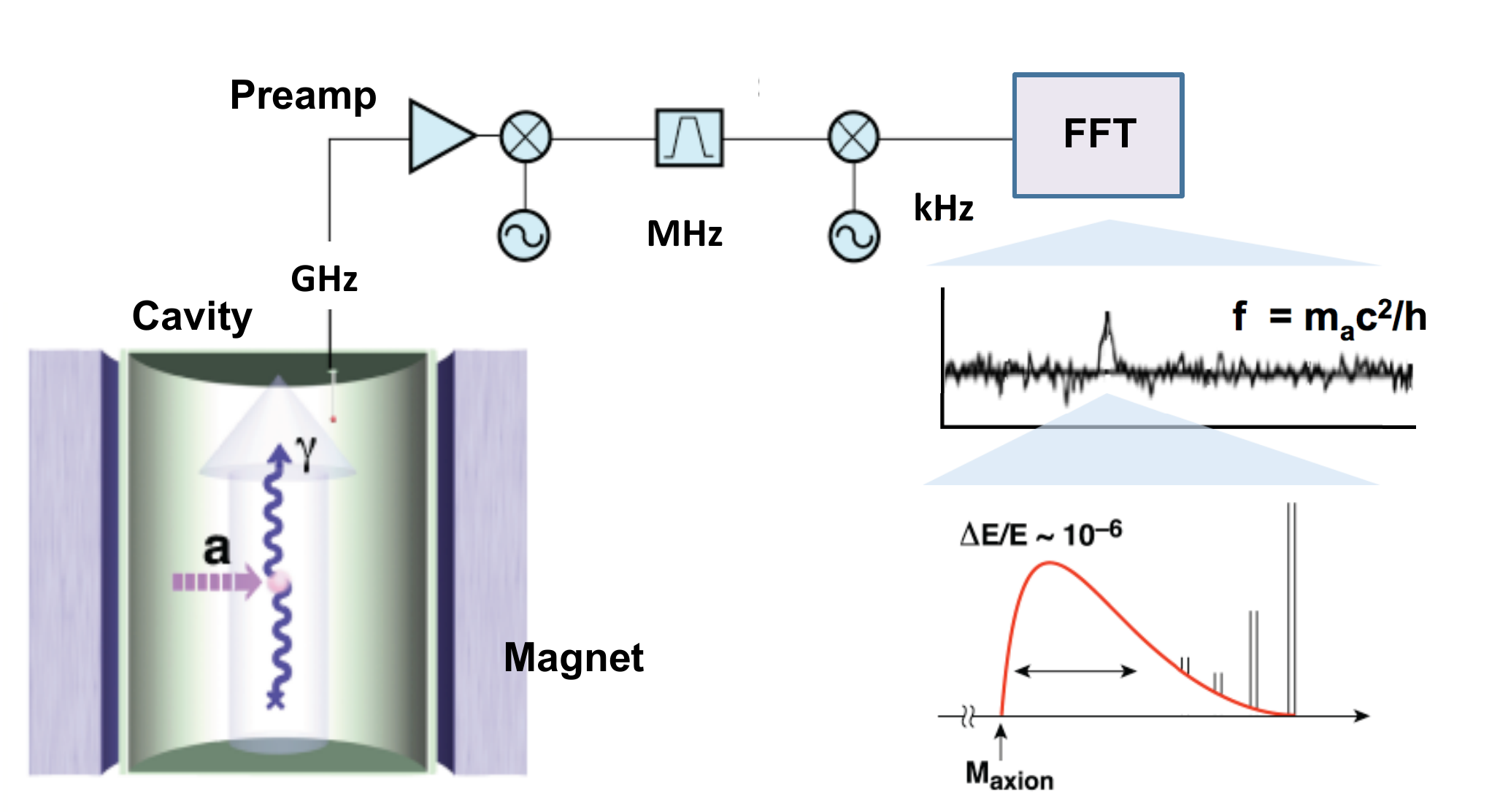}}
%\centerline{\includegraphics[width=12cm]{Fig_1.eps}}
\caption{
Schematic of the microwave cavity search for dark-matter axions.  Axions resonantly convert to a quasi-monochromatic microwave signal in a high-$Q$ cavity in a strong magnetic field; the signal is extracted from the cavity by an antenna, amplified, downconverted, and the power spectrum calculated by a FFT. Any fine structure in the signal would reveal important information about the formation of our galaxy.
\label{fig:1}}
\end{figure}

\noindent Axions constituting our galactic halo may be detected by their resonant conversion to photons in a microwave cavity permeated by a magnetic field,\cite{8} the conversion power being

\begin{equation}
P = g^2_{a\gamma\gamma}\left(\frac{\rho_a}{m_a}\right)B^2_0 V C \mbox{ min}(Q_L,Q_A)\,.
\label{equ:2}
\end{equation}
\\ \noindent 
The physics parameters, beyond the control of the experimentalist, are the axion-photon coupling constant $g_{a\gamma\gamma}$, the axion mass $m_a$, and the local density of axions in the halo, $\rho_a$.  Within experimental control are the magnetic field strength $B_0$,  the volume of the cavity $V$, the mode-dependent form-factor $C$, which measures the overlap of the electric field of the cavity mode with the external magnetic field, and the loaded quality factor $Q_L$ of the cavity, assumed to be critically coupled. The resonant conversion condition is that the frequency of the cavity must equal the mass of the axion\footnote{A frequency of 1 GHz corresponds to 4.136 $\mu$eV.}, $h\nu = m_a c^2[1+\frac{1}{2}½O(\beta^2)]$, where $\beta \approx 10^{-3}$ is the galactic virial velocity. The signal is thus monochromatic to 10$^{-6}$, which implies $Q_A\sim10^{6}$.  The search is performed by tuning the cavity in small overlapping steps (Fig.~\ref{fig:1}).  \\

\noindent The axion-photon coupling constant $g_{a\gamma\gamma}$ is rather insensitive to the choice of axion model, differing only by factors of order $\sim 7$ for various models which have been investigated to date.  Representative couplings from two different classes of models, denoted KSVZ\cite{9,10} and DFSZ,\cite{11,12} have historically served as goals for experiments.  The axion-photon coupling constant $g_{a\gamma\gamma}$ has dimension GeV$^{-1}$ and is proportional to the axion mass. Different classes of axion models may instead be characterized in a mass-independent way by a dimensionless parameter $g_\gamma \propto g_{a\gamma\gamma}: g_\gamma=-0.97$ for KSVZ axions and +0.36 for DFSZ axions.    \\

\noindent The sensitivity of the experiment is governed by the Dicke radiometer equation\cite{13}
\begin{equation}
\frac{S}{N} = \frac{P_{SIG}}{k T_{SYS}}\sqrt{\frac{t}{\Delta\nu}}\ ,
\label{equ:3}
\end{equation}
where $S/N$ is the signal-to-noise ratio, and the system noise temperature $T_{SYS} = T + T_N$  is the sum of the physical temperature $T$ and the intrinsic amplifier noise temperature $T_N$, with $k$ the Boltzmann constant.  The integration time is $t$, and the bandwidth of the axion signal is $\Delta\nu$, where it is assumed that the resolution of the spectral receiver is much better than the width of the axion signal. This condition will be important later when discussing the potential application of single-photon detection schemes to microwave cavity experiments. \\

\begin{figure}[b]
\centerline{\includegraphics[width=12.7cm]{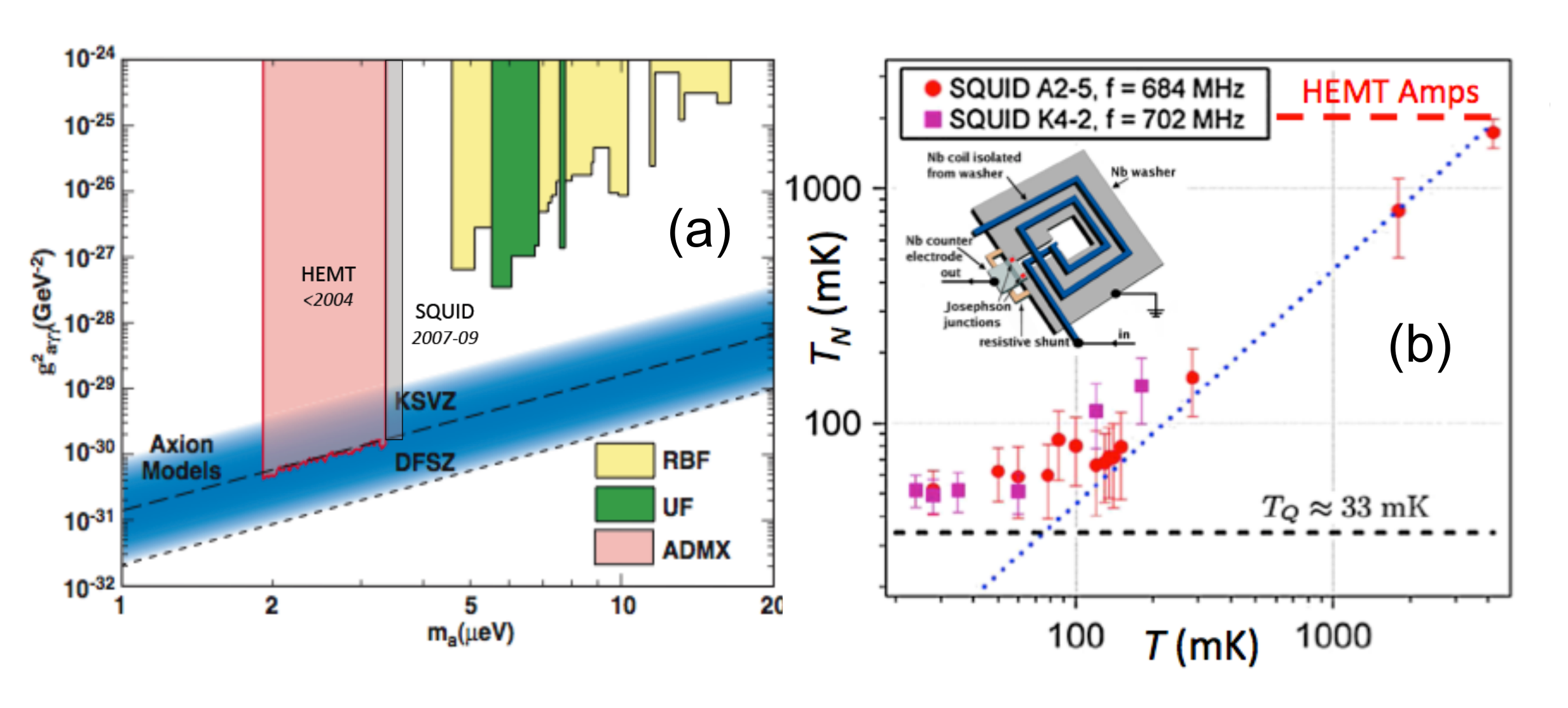}}
%\centerline{\includegraphics[angle=90,width=14.2cm]{IJMPA_Fig_2_crop.eps}}
%\centerline{\includegraphics[width=12cm]{Fig_2.eps}}
\caption{(a) The 90\% c.l. exclusion regions for the RBF,\cite{14,15} Florida\cite{16} and ADMX\cite{17} experiments. (b) The Microstrip-coupled SQUID Amplifier (MSA), showing the intrinsic noise temperature $T_N$ vs. physical temperature $T$; the MSA exhibits near-quantum-limited performance at dilution refrigerator temperatures.\cite{18,19,20}
\label{fig:2}}
\end{figure}

\noindent Two pilot efforts, at Brookhaven\cite{14,15} and the University of Florida,\cite{16} were mounted within a few years of the original experimental concept.\cite{8}  As the best conventional amplifiers at that time (e.g. HEMT amplifiers) had noise temperatures in the $T_N \sim 3-20$ K range, they did not have the sensitivity to reach Peccei-Quinn axions (Fig.~\ref{fig:2}a), but developed much of the design and know-how current experiments still build on. \\

\noindent The Axion Dark Matter eXperiment (ADMX) marked both a significant scale-up of the volume of the microwave cavity and a reduction in system noise temperature, excluding KSVZ axions over the mass range $m_a = 1.9-3.6\;\mu$eV.\cite{17}  In its most recent phase, the experiment utilized Microstrip-coupled SQUID Amplifiers (MSA), which have demonstrated near-quantum-limited noise temperatures ($k T_N = h\nu$) on the bench when cooled to $T < 100$ mK.\cite{18,19,20} ADMX will soon be upgraded with a dilution refrigerator that will enable the full capability of the MSA to be realized in operation. For a more detailed overview of axion theory and experimental searches, see S. J. Asztalos \emph{et. al.}\cite{35}

\section{ A strategy for higher-mass axions}
The microwave cavity experiment has made significant progress in advancing sensitivity to axions of a few $\mu$eV mass, or around 1 GHz. Extending the search to higher frequencies while maintaining comparable sensitivity requires development of new resonator and amplifier technologies for the $>$10 GHz range. 
\subsection{ADMX-HF}
\begin{figure}[b]
\centerline{\includegraphics[width=12cm]{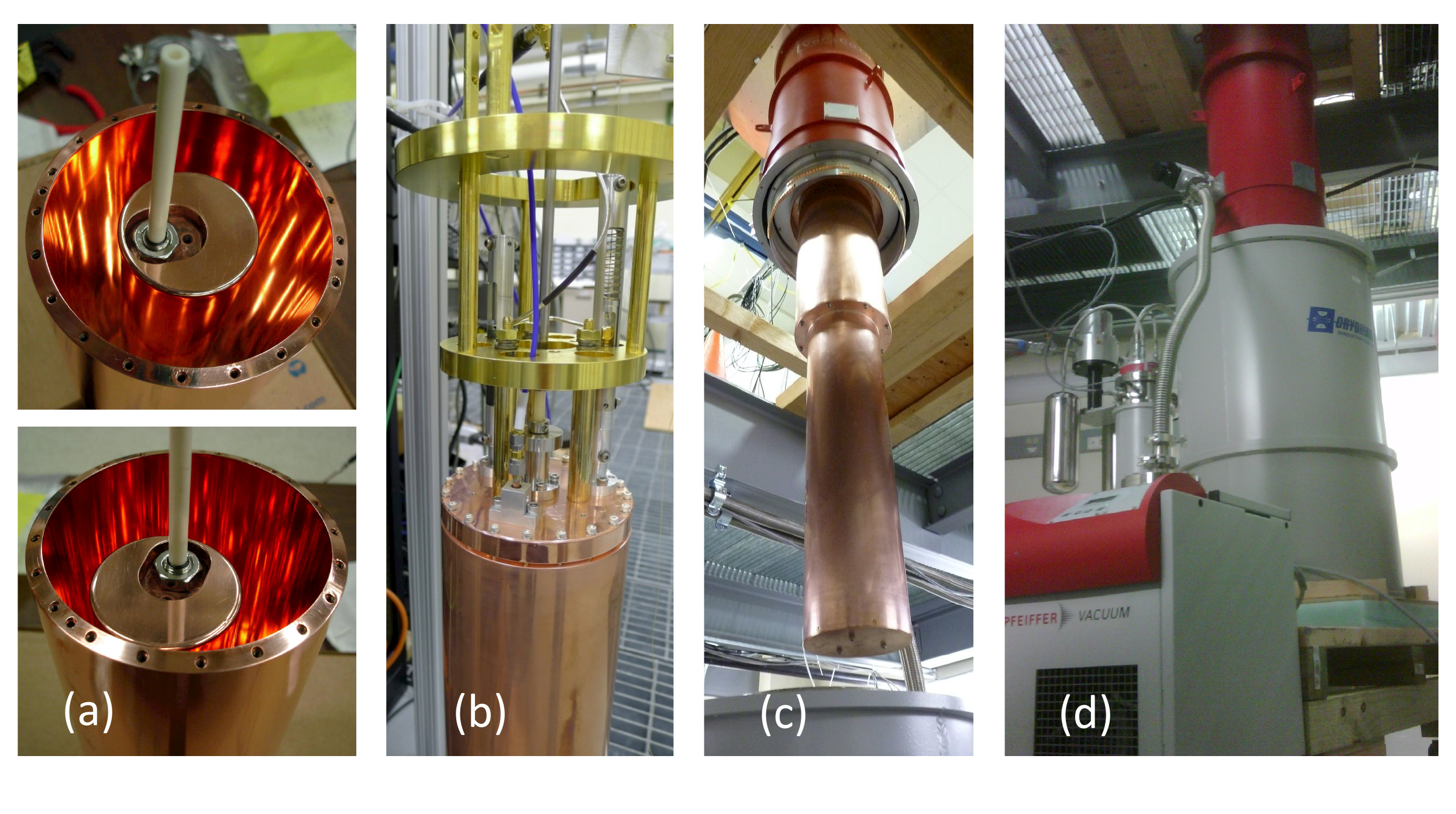}}
%\centerline{\includegraphics[angle =90,width=14.2cm]{IJMPA_Fig_3.eps}}
\caption{(a)  Tuning of single-rod cavity.  (b)  Microwave cavity mounted on the gantry that thermally couples it to the mixing chamber of the dilution refrigerator.  (c)  Gantry with thermal shields suspected from the dilution refrigerator being lowered into the magnet.  (d) Fully-integrated experiment.}
\label{fig:3}
\end{figure}

The Axion Dark Matter eXperiment at High Frequencies (ADMX-HF),\footnote{ADMX-HF is a collaboration of Yale, UC Berkeley, JILA/Colorado, and LLNL.} now entering its commissioning phase at Yale, was specifically designed both as a \emph{pathfinder} to take first data in the 5 -- 25 GHz range ($\sim20-100\ \mu$eV) of the axion model band, and as an \emph{innovation test-bed} for developing new concepts and technologies for the microwave cavity experiment.  The magnet is a relatively small high-field, high-aspect ratio superconducting solenoid (9 T, 40 cm $\times$ 16.5 cm clear bore), designed to have a very small radial component of the field on the surface of the microwave cavity (B$_r$ $<$ 50 G), enabling the exploration of hybrid superconducting cavities, discussed below.  Initially, the cavities (25 cm $\times$ 10 cm I.D.) will consist of a thin layer of copper electrodeposited on stainless steel and annealed.  Tuning will be accomplished via the radial displacement of a metal rod or via a combination of moving and stationary posts.  The experiment will operate at dilution refrigerator temperatures (T $\sim$ 25 mK).  Figure~\ref{fig:3} shows the experiment in various stages of integration.\\

\noindent ADMX-HF will use Josephson Parametric Amplifiers (JPAs) as the cryogenic preamplifier from the outset; as described in further detail below, they readily achieve quantum-limited noise performance and are tunable over an octave in frequency.   Such low noise is critical as the expected signal is extraordinary small, a few yoctowatts. \\

\begin{figure}[b]
\centerline{\includegraphics[width=13.5cm]{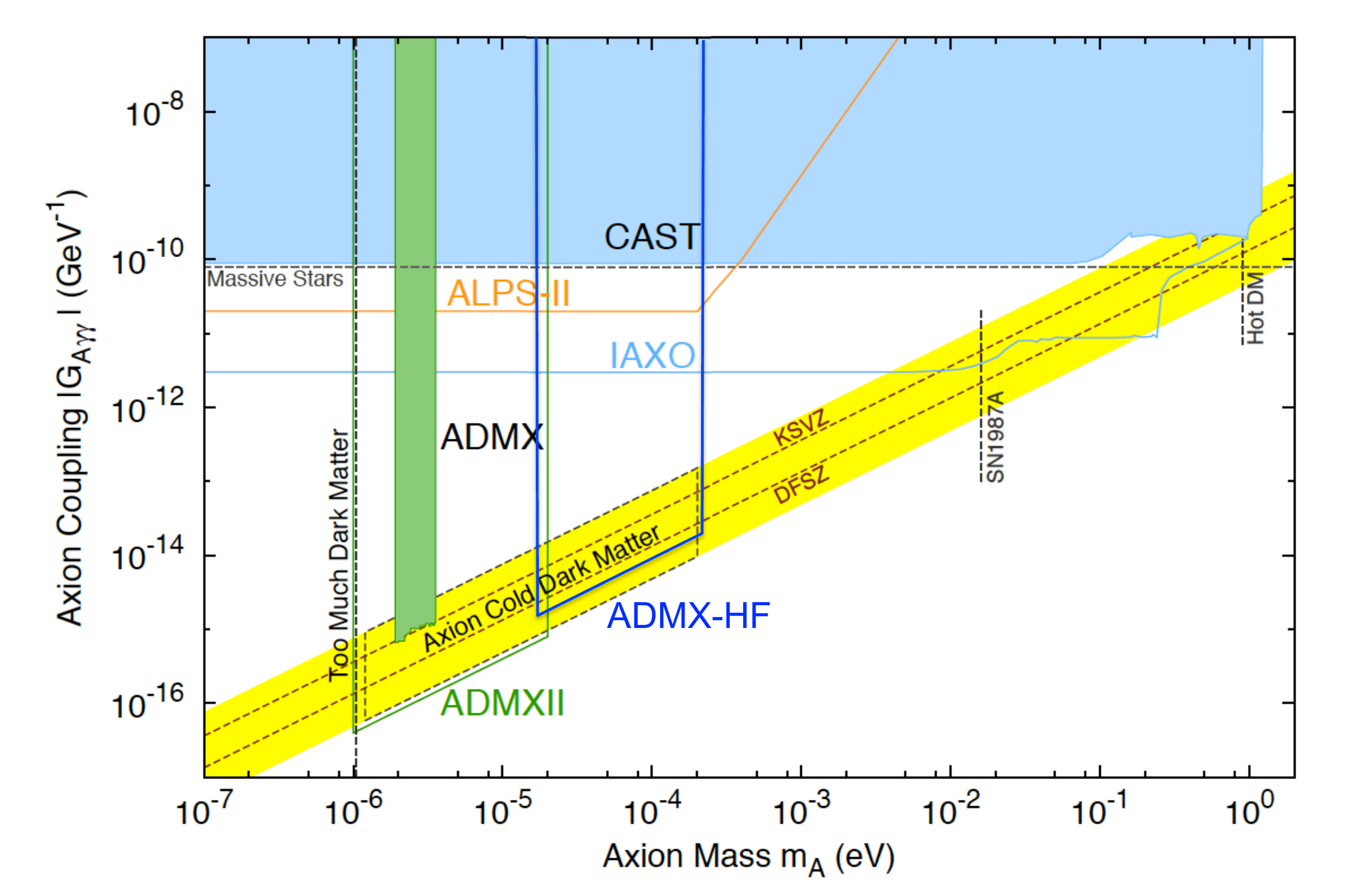}}
%\centerline{\includegraphics[angle =90,width=14.2cm]{IJMPA_Fig_4.eps}}
\caption{The mass and coupling parameter space accessible within five years after full development of ADMX and ADMX-HF.  The yellow band notionally represents the expected axion model region; shown also are the existing (CAST)\cite{31} and proposed (IAXO)\cite{32,33} direct solar limits, and limits achievable from the ALPS-II\cite{34} resonant photon regeneration experiment.  }
\label{fig:4}
\end{figure}

\noindent Together ADMX and ADMX-HF represent a coordinated US dark-matter axion search, with the goal of covering the space in ($m_a$, $g_{a\gamma\gamma}$)  represented in Figure~\ref{fig:4}.

\subsection{New paradigms for the microwave cavity experiment at high frequencies }
Scaling relationships are useful in developing experimental strategy.  For a fixed scan rate, the smallest axion-photon coupling, g$_\gamma$, that can be detected depends on the experimentally controlled parameters as:
\begin{equation}
g_\gamma \propto T^{1/2}_{SYS}B^{-1}V^{-1/2}Q_L^{-1/4}t^{-1/4}.
\end{equation}
Similarly, fixing the coupling $g_\gamma$  for which axions would be detected or excluded, the mass scan rate scales as:
\begin{equation}
\frac{d\nu}{dt} \propto B^4V^2\frac{Q_L}{T_{SYS}}.\\
\end{equation}
$Q_L$ is the loaded quality factor of the cavity, assumed to be critically coupled, i.e.  $Q_L = Q / 2$. \\

\noindent Extending the experiment to much higher masses while remaining sensitive to Peccei-Quinn axions is challenging.  There are practical limits to the size and strength of the magnet, and significantly increasing the integration time is not feasible given the decades of mass to be scanned. Furthermore, there are implicit frequency dependencies of both the cavity quality factor ($Q \propto \nu ^{-2/3}$) for copper cavities, and the standard quantum limit for the system noise temperature ($T_{SYS} \propto \nu $), which conspire against a definitive experiment. \\

\noindent This suggests two lines of R$\&$D, being pursued by the collaboration:
\begin{itemize}
\item Hybrid superconducting cavities, i.e. cavities with thin-film depositions of a Type-II superconductor on all cylindrical surfaces to boost $Q$ by an order of magnitude.\\
\item Operation of JPAs in squeezed-state mode to evade the standard quantum limit;  single-photon detection based on superconducting qubits will also be explored.\\
\end{itemize}

\begin{figure}[b]
\centerline{\includegraphics[width=9cm]{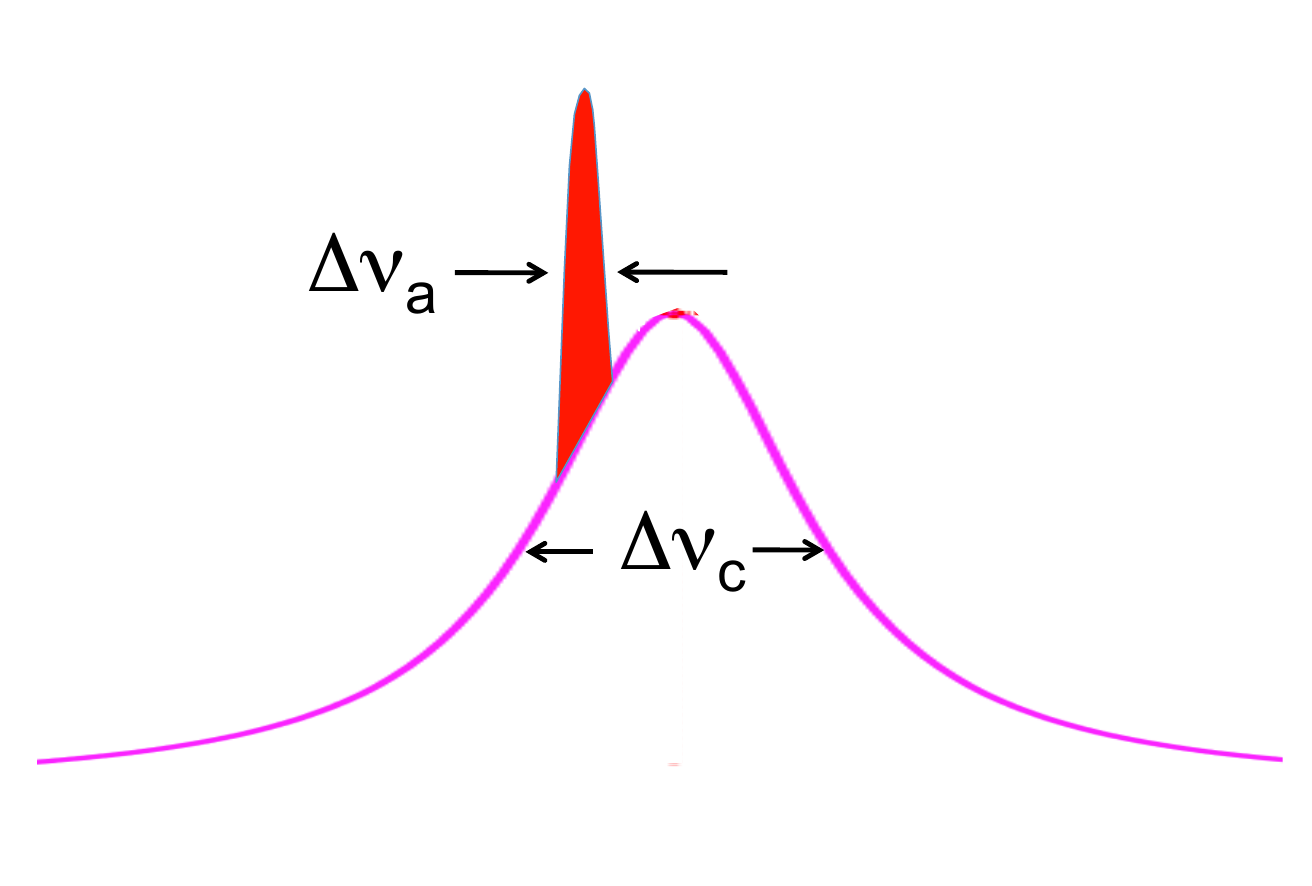}}
%\centerline{\includegraphics[width=12cm]{IJMPA_Fig_5-1.pdf}}
%\centerline{\includegraphics[angle =90,width=14.2cm]{IJMPA_Fig_5.eps}}
\caption{Schematic representation of the virialized axion signal, $Q_a =E_a/\Delta E_a \sim 10^6$, on top of the Lorenzian bandpass of the cavity, with loaded quality factor, $Q_L = \nu/\Delta \nu$.  }
\label{fig:5}
\end{figure}

\noindent While implementing either of these concepts will improve the sensitivity of ADMX-HF at the present frequency of 5 GHz, if both can be achieved, a viable strategy for high masses becomes possible. We recently analyzed the trade-offs between linear amplifiers and single-photon detectors for the microwave cavity experiment and found that, if the cavity quality factor can be made comparable to that of the axion signal, i.e. $Q_L/Q_a \sim 1/3$, the cross-over point where bolometric detection becomes favored is surprisingly low, around 10 GHz.\cite{21}  It is important to note that the cavity $Q_L$ enters in as noise power accrues from the entire cavity bandwidth once spectral resolution is forsaken; see Figure~\ref{fig:5}.  

\section{Technology development for a definitive high mass axion experiment}
The underlying physics that explains the path from linear amplifiers to squeezed states of the vacuum and finally to true single-photon detectors, and realistic projections for the ultimate reach of the microwave cavity experiment in ($m_a, g_{a\gamma\gamma}$), will be developed in detail in a forthcoming publication.  We restrict ourselves below to a first look at the state of the technology of hybrid superconducting cavities and Josephson parametric amplifiers.

\subsection{Hybrid superconducting cavities}
The microwave cavity experiment puts a series of unusual constraints on microwave cavity design.  The resonator must have a good form factor C: this condition forces experiments to use TM$_{010}$-like modes. The resonator must also have as high a $Q$ as possible, extreme longitudinal uniformity to avoid mode-localization, a reasonable aspect ratio to minimize the number of avoided crossings with intruder TE-modes, and a large dynamic tuning range ($\pm30\%$).  At frequencies of $\sim$1 GHz, experiments have used a single copper cylindrical cavity, tuned by radially displacing one or more metal or dielectric rods inside the cavity.  \\

\begin{figure}[b]
\centerline{\includegraphics[width=12cm]{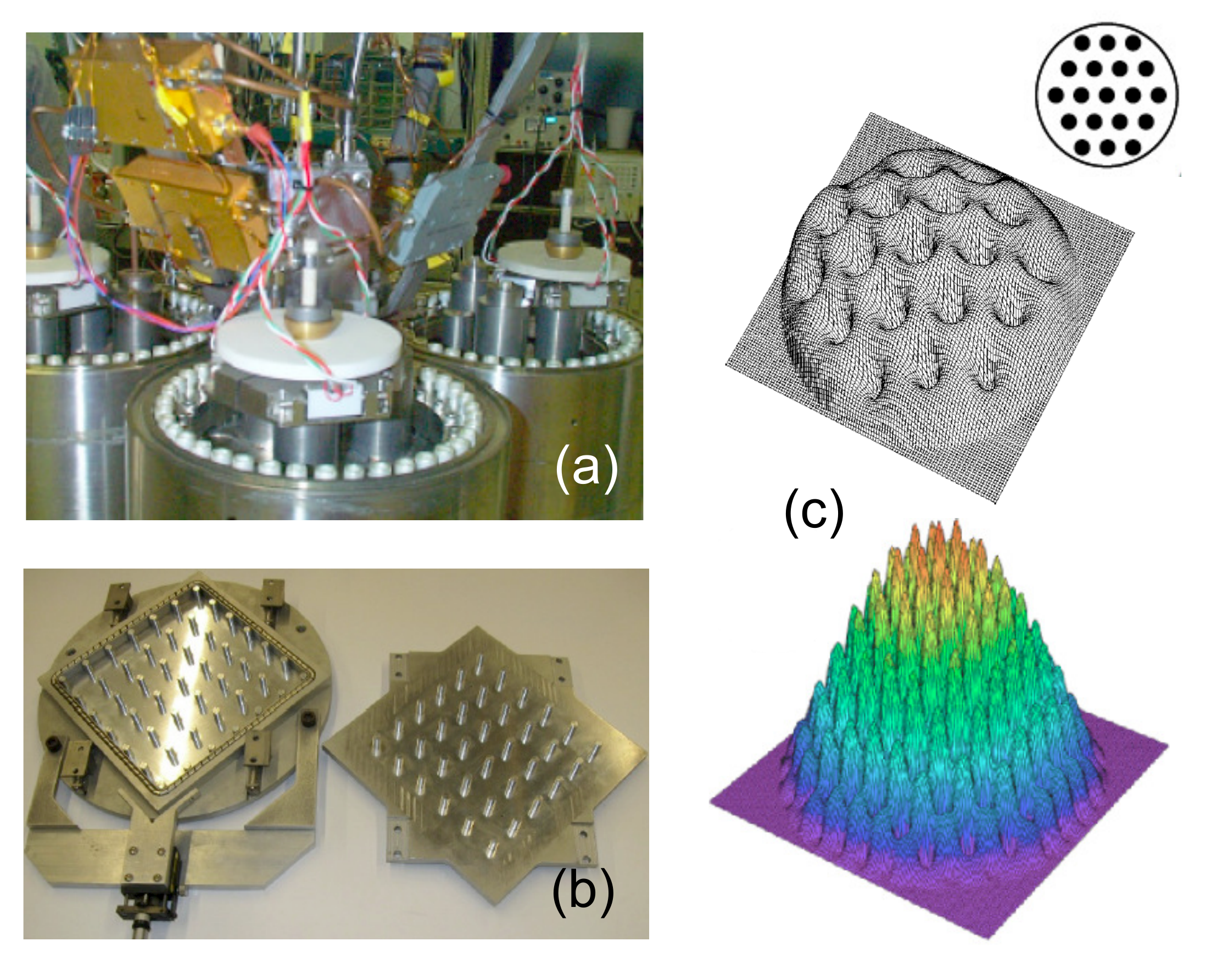}}
%\centerline{\includegraphics[angle =90,width=14.2cm]{IJMPA_Fig_6.eps}}
\caption{(a) Four-cavity array used in ADMX.\cite{22} (b) Prototype of a photonic band-gap resonator in the multi-GHz range.  (c)  Calculated E-field for the TM$_{010}$ mode of a photonic band-gap resonator with 19 and 91 posts. }
\label{fig:6}
\end{figure}

\noindent A significant challenge for the experiment is the development of resonators for much higher frequencies, i.e. $>$ 10 GHz.  By definition, the characteristic size of such structures goes inversely with the frequency; at the same time however, one wishes to utilize as much available volume within the magnet as possible to maximize signal power. Extensive modeling and prototyping of resonator concepts has been carried out for many years, including power-combining multiple cavities \cite{22} as well as photonic band-gap cavities, i.e. cavities loaded with lattices of metallic posts (Fig. ~\ref{fig:6}).\\

\noindent Photonic band-gap cavities have been studied by the accelerator physics community for an X-band linear collider \cite{23} and represent an attractive option for ADMX-HF.  A tuning scheme with good dynamic range which preserves lattice symmetry has been conceived.\\

\noindent The quality factor of a cavity is given generically by 
\begin{equation}
Q=\left( \frac{V}{S\cdot\delta}\right)\times(\mbox{geometrical factor}),
\end{equation}
where $V, S$ are the volume and surface area of the cavity respectively, $\delta$ is the skin depth, and the geometrical factor is mode-dependent, but of order unity.  For pure copper at cryogenic temperatures, the microwave conductivity is limited by the anomalous skin depth: $\delta_{anom}  =  2.8 \times 10^{-5}$ cm ($\nu$ [GHz])$^{-1/3}$.  The first ADMX-HF copper cavity, critically coupled to its antenna, will have $Q_L$ = 30,000, a factor of $\sim$30 lower than the $Q$ of the virialized axion signal.  The $Q$ will deteriorate further as the volume-to-surface ratio at higher frequencies.  \\

\begin{figure}[b]
\centerline{\includegraphics[width=15cm]{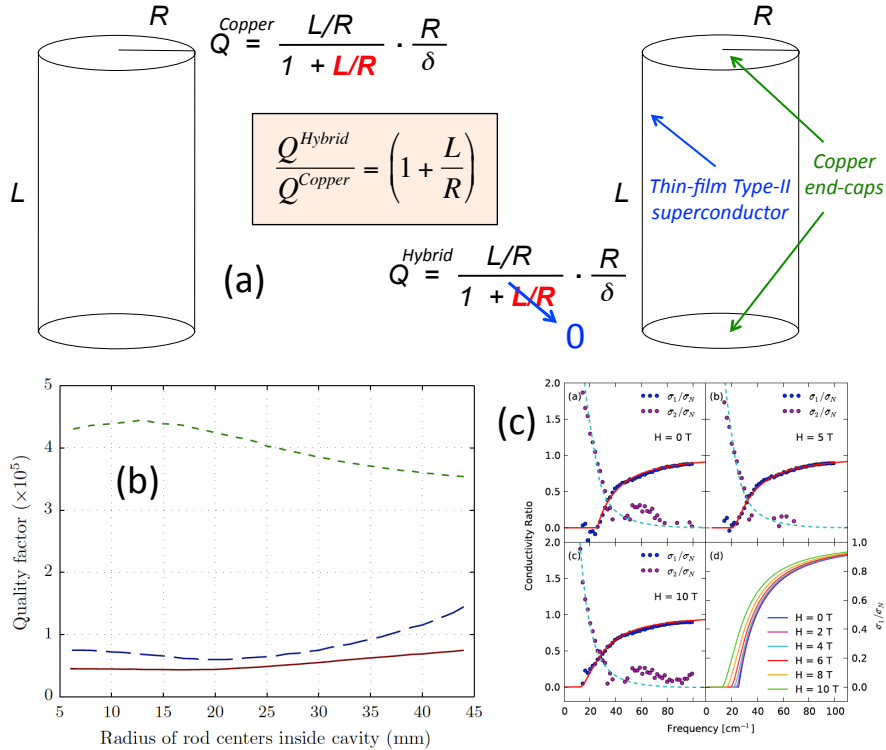}}
%\centerline{\includegraphics[angle =90,width=14.2cm]{IJMPA_Fig_7.eps}}
\caption{(a)  For an empty cavity, $Q$ of the TM$_{010}$ mode improves by a factor of (1+L/R) when the barrel is coated with a thin-film superconductor.  (b)  Importance of coating all cylindrical surfaces, including the tuning rods:  All copper (solid line); barrel coated only (long dash); both barrel and tuning rod (short dash).  Example for a single-rod cavity; dimensions for cavity currently used in ADMX-HF.  (c)  Demonstration of perfect microwave reflectivity for  B$_\parallel$  = 10T in NbTiN thin films; Xi et al. \cite{24} }
\label{fig:7}
\end{figure}

\noindent However, if one could make a cavity in which all surfaces parallel to the magnetic field (i.e., the barrel and tuning rods) were superconducting, the $Q$ would increase by $O$(10) depending on the exact aspect ratio of our cavity, making the power and signal-to-noise equivalent to that of an experiment an order of magnitude larger (Fig.~\ref{fig:7}a,b).\\

\noindent Recent measurements of the optical properties of thin film Type-II superconductors at the NSLS have shown that Nb$_{0.5}$Ti$_{0.5}$N films of 10 nm thickness support B$_\parallel$ = 10 T, and are perfectly microwave reflecting up to $>$100 GHz (Fig. ~\ref{fig:7}c).\cite{24}  The deposition and characterization of thin film superconductors has been thoroughly studied; in particular, Nb$_x$Ti$_{1-x}$N is easy to fabricate, relatively non-critical stoichiometrically, and makes a durable and long-lasting film.  RF losses in superconducting cavities result from flux vortices, and thus the film thickness should not exceed the characteristic spacing of the vortices to minimize the number parallel to the surface.  The ADMX-HF magnet was designed with high field uniformity (B$_r \ <$ 50 G) to reduce flux vortices perpendicular to the surface to an acceptable level, thus enabling R$\&$D with hybrid superconducting cavities.\\

\begin{figure}[b]
\centerline{\includegraphics[width=12cm]{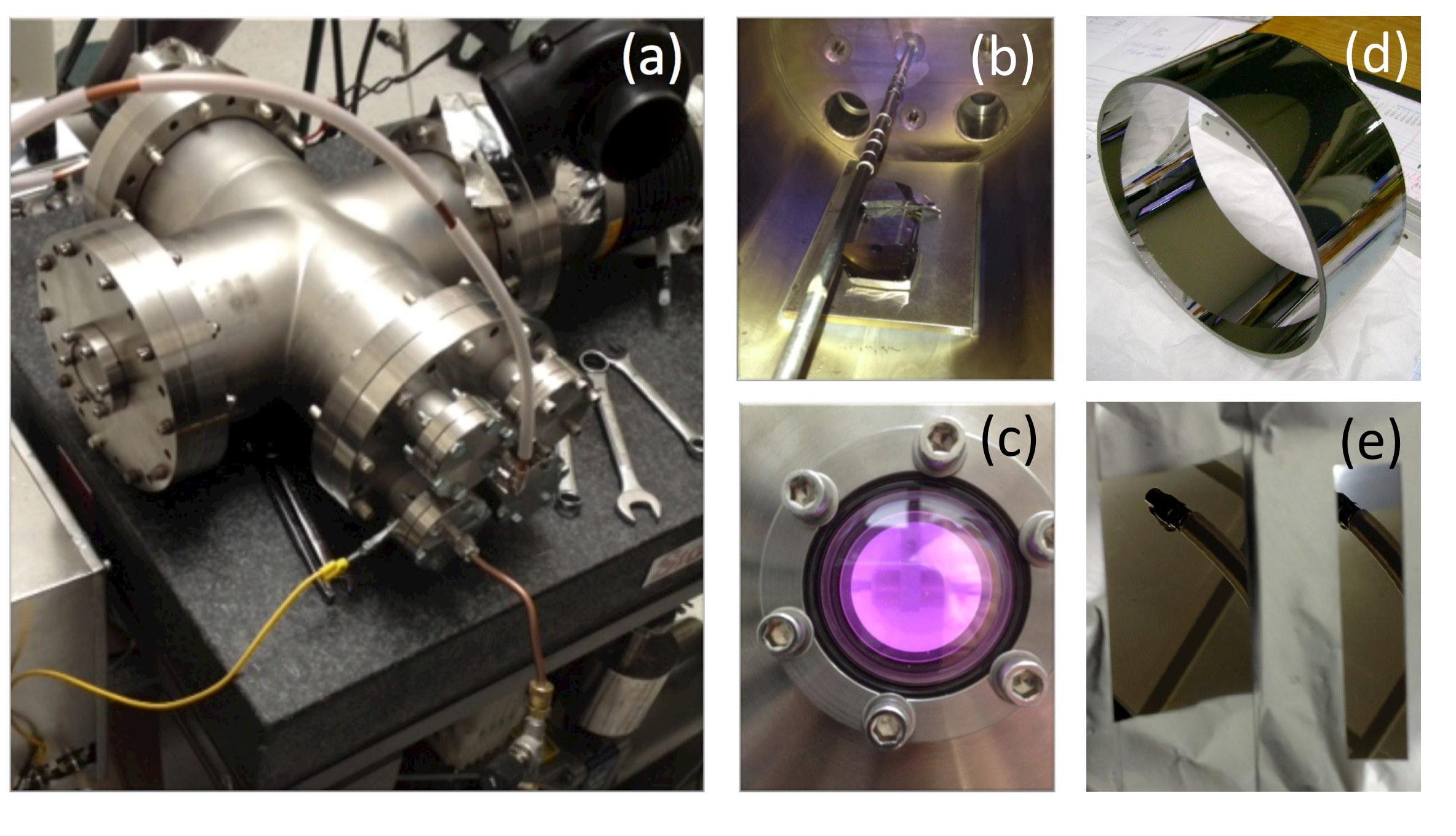}}
%\centerline{\includegraphics[angle =90,width=14.2cm]{IJMPA_Fig_8.eps}}
\caption{(a)  RF plasma deposition chamber.     (b) Chamber interior, showing the central linear antenna, a water-cooled tube of Nb with Ti sleeving.  (c) Plasma discharge of Ar + N.  Uniform mirror-like coatings have been made on glass and silicon substrates for testing with (d) cylindrical interiors, and (e) planar coupons. }
\label{fig:8}
\end{figure}

\begin{figure}[b]
\centerline{\includegraphics[width=13.1cm]{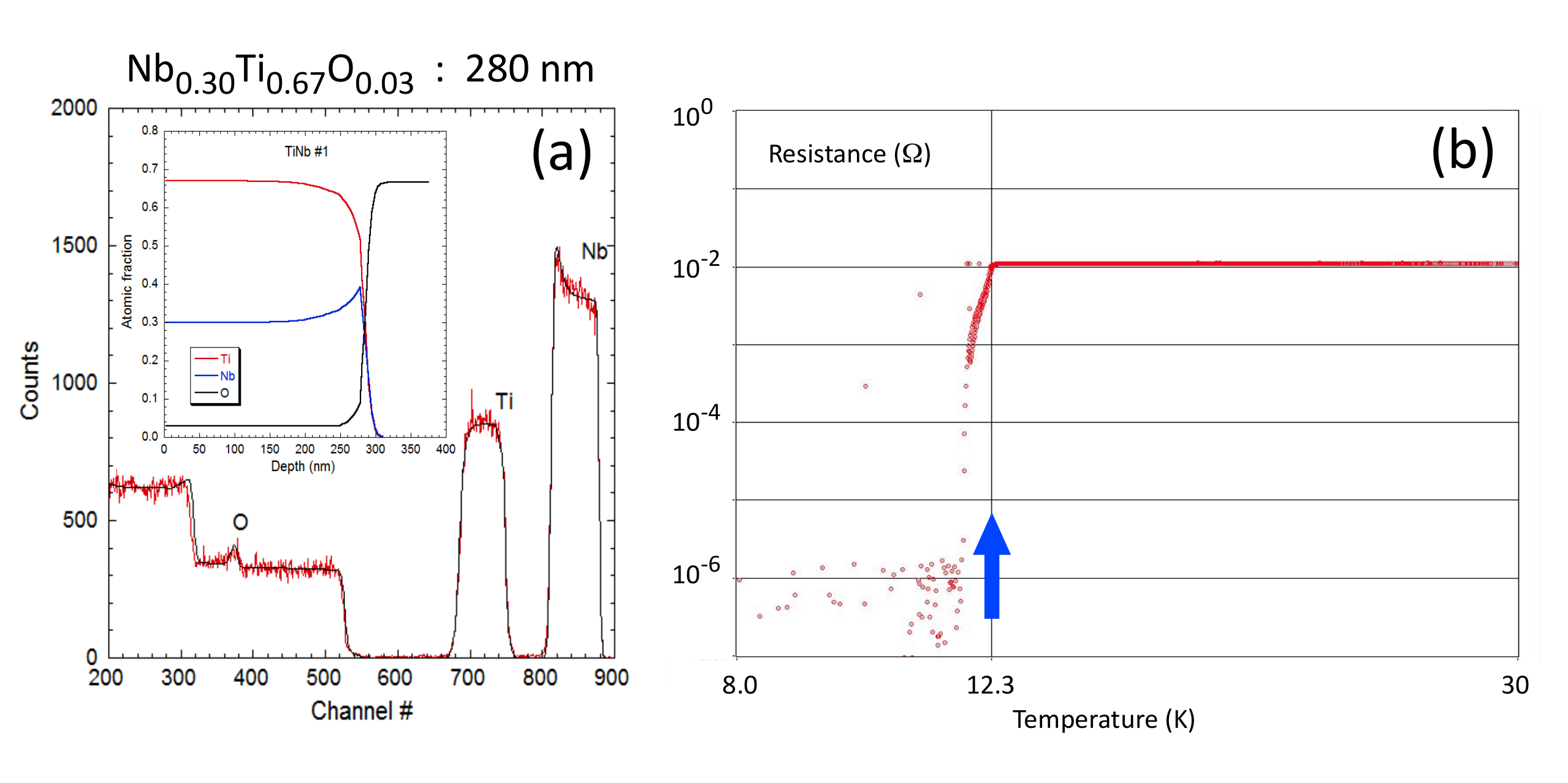}}
%\centerline{\includegraphics[angle =90,width=14.2cm]{IJMPA_Fig_9.eps}}
\caption{(a) Rutherford backscattering profile of a NbTiN thin film, indicating well-controlled stoichiometry and negligible oxygen contamination.  (b) DC resistance measurement of a thin film showing near-maximum $T_C$.}
\label{fig:9}
\end{figure}

\noindent Thin films of Nb$_x$Ti$_{1-x}$N have been produced on planar glass and silicon substrates by RF plasma deposition by both Berkeley-LLNL and Yale groups, and ultimately on the interior of quartz tubes (Fig.~\ref{fig:8}); these films possess good uniformity, approximately the desired composition, and low oxygen content.  Most importantly, films readily exhibited DC superconductivity in 4-wire LHe dunk tests, exhibiting $T_c\approx$ 12 K (Fig.~\ref{fig:9}).  Tests of RF superconductivity in prototype 10 GHz cavities are planned for the near future; if successful, a hybrid superconducting cavity could be introduced into ADMX-HF early in its data-taking program.\\

\subsection{Josephson parametric amplifiers, squeezed states and single quantum detection}
In the microwave cavity axion search, the measurement problem is one of searching for a feeble microwave signal of unknown frequency. The detection time is greatly increased by noise introduced by the microwave measurement apparatus itself. In many cases, microwave measurements must add noise of a quantum origin, and if the dominant noise source is this quantum noise, the apparatus is said to be Òquantum-limited.Ó \\

\noindent In the past five years, microwave measurement technology has advanced enormously. Quantum-limited measurements at microwave frequencies, once considered a remote goal, have become routine within the community of superconducting quantum information processing. Specifically, at frequencies between 4 -- 8 GHz, a new generation of amplifiers based on superconducting Josephson junction physics operate almost as mathematical idealizations of amplification. These Josephson parametric amplifiers (JPAs) add only the noise required by quantum mechanics of any linear amplifier that preserves the phase of the amplified signal.  In 2007, the best available 4 -- 8 GHz amplifiers used transistors and added noise about 30 times the quantum-limited noise, thus requiring $\sim$1000 times longer to detect an axion signal than a quantum-limited JPA.  For the foreseeable future, JPAs will be the critical measurement technology enabling the microwave cavity axion search above 4 GHz. \\

\noindent At even higher frequencies, the quantum noise itself becomes unacceptably large compared to the putative axion signal. Fortunately, JPAs and related superconducting devices allow one to circumvent even this fundamental source of noise.  A seminal discussion of the relationship between non-commuting observables and quantum noise is given by Caves.\cite{25} In short, JPAs have two special properties, both of which are required to overcome quantum noise. First, they can circumvent the quantum limit on amplification by operating in a ``phase-sensitive mode." Second, they can squeeze the quantum fluctuations of the microwave vacuum.\\

\noindent JPAs derive their remarkable properties from the circuit element known as a Josephson junction (JJ). For sufficiently small currents, a JJ is a two terminal circuit element that behaves like an electrically non-linear inductor. The inductance of the element is a function of the current flowing through it. Importantly, the dissipation of this non-linear inductor is vanishingly small; it is a supercurrent that flows through the element.  Low-dissipation and strong non-linearity can be combined to build either a quantum-limited amplifier or a noiseless amplifier. By connecting the JJ to a capacitor, one realizes a non-linear resonant circuit, one whose resonant frequency depends on the current in the circuit itself. A `pump' tone can cause the circuit's resonance to vary at twice its average frequency. Any current flowing in the circuit when the pump is turned on will grow if it is in phase with the pump, but shrink if it is 90 degrees out of phase. This phenomenon is the electrical analog of a familiar mechanical parametric amplifier, i.e. a child's swing. \\

\noindent In order to create an amplifier based on this parametric process, the non-linear circuit is weakly coupled to a transmission line, which will form both the input and output of the amplifier. Current fluctuations are introduced into the circuit by microwave fields propagating in the transmission line towards the circuit. Those fluctuations enter the resonant circuit where they are amplified or deamplified and then emerge as a wave traveling away from the circuit. In steady state, this arrangement is an amplifier where the signal to be amplified is incident on the circuit and the amplified signal is the signal reflected from that circuit. If the signal has a fixed phase relationship with the pump, the JPA acts as a phase sensitive amplifier; otherwise it is phase-insensitive. \\

\noindent There are two important operational constraints on JPAs.  First, they are naturally narrow-band amplifiers, only capable of amplifying signals in a narrow frequency range near the non-linear circuit's resonance frequency. Second, in order to provide a quantum-limited microwave measurement they must also operate with sufficient gain to overwhelm any noise introduced by subsequent stages of amplification. In principle, JPAs can operate with arbitrarily large gain. As the pump power increases, the gain increases, eventually diverging at a critical power; near this critical power however, the gain is also arbitrarily sensitive to any source of interference. \\

\begin{figure}[b]
\centerline{\includegraphics[width=12cm]{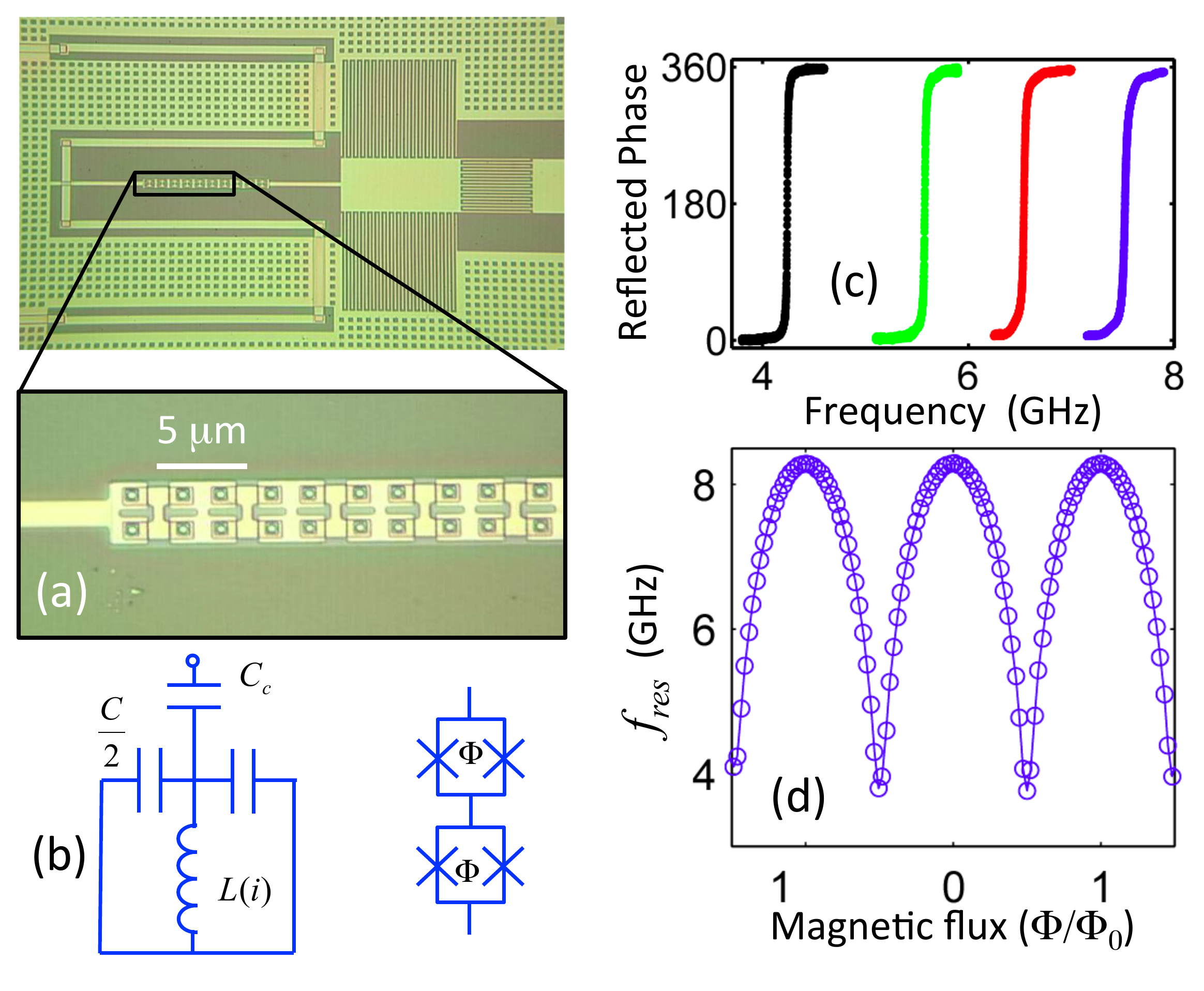}}
%\centerline{\includegraphics[angle =90,width=14.2cm]{IJMPA_Fig_10.eps}}
\caption{Tunable Kerr Circuit.  (a) Microphotograph of recent TKC design with LC circuit made from niobium (light green) on a silicon substrate. Insert: Non-linear and flux-tunable inductance made from an array of SQUIDs in series.  (b) TKC circuit diagram, with current-dependent inductance. (c) The phase of a microwave tone reflected off of the input capacitance $C_c$ indicates a resonance by a spectrally sharp phase wrap. Colors correspond to different applied fluxes. (d) Resonance frequency tuned with applied flux over an octave.}
\label{fig:10}
\end{figure}

\begin{figure}[b]
\centerline{\includegraphics[width=12cm]{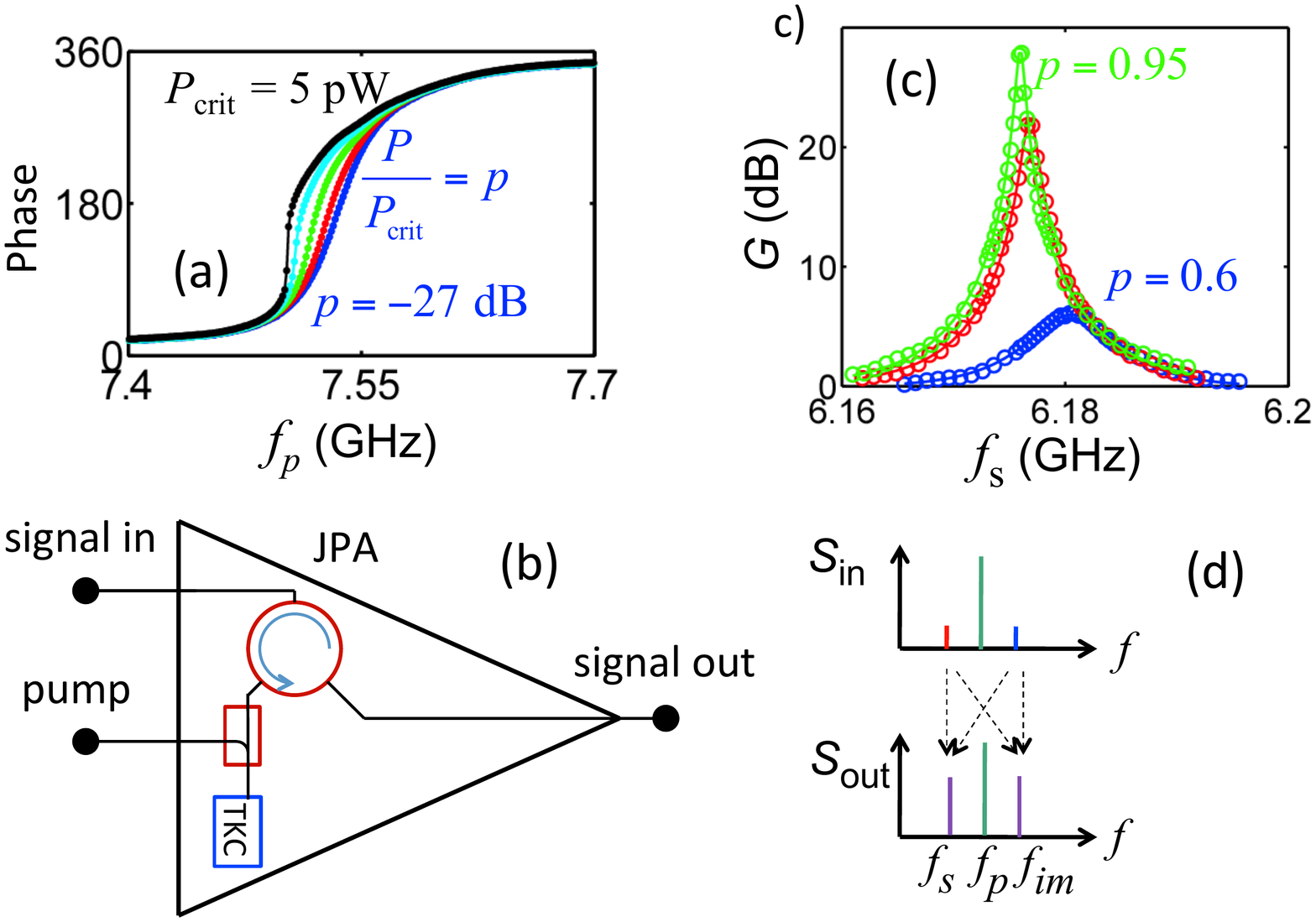}}
%\centerline{\includegraphics[angle =90,width=14.2cm]{IJMPA_Fig_11.eps}}
\caption{JPA Operation. (a) The reflected phase from the TKC is strongly power-dependent, the phenomenon from which gain is derived. (b) A JPA is formed from a TKC, a directional coupler that adds a pump tone to the input signal, and a circulator that separates the amplified output from the input. (c) Gain versus signal frequency at different values of pump power, showing the increase in gain as the pump power approaches critical power. (d) A spectral domain diagram shows that either phase sensitive or phase insensitive amplification is possible using a JPA. If the signal is represented by the red tone at the input, at the amplifier output it appears at the same frequency, but the tone is a superposition of the signal and noise that enters at the idler frequency $f_{im} = 2f_p - f_s$. On the other hand, if the signal were regarded as both the red and blue tones, as would be the case if the signal of interest were the amplitude modulation in a copy of the pump tone, the amplification would be noiseless.}
\label{fig:11}
\end{figure}

\noindent In 2007 and 2008, one of us (Lehnert) introduced a new JPA design that allowed it to be operated with large gain and with a tunable band, overcoming several limitations of earlier designs. Specifically, the new design isolated the JJ from all signals except the pump and the input signal by embedding it in a single mode resonant circuit (Fig.~\ref{fig:10}). The pump is applied near the circuit's resonance. When exciting the JJ's non-linear circuit, the pump sharpens the circuit's phase response and experiences a power-dependent phase shift upon reflection, behavior analogous to optical Kerr cavities (Fig.~\ref{fig:11}), and the non-linear and tunable resonant circuits will be referred to here as tunable Kerr circuits (TKC). The addition of a small signal to the pump creates a large change in the phase of the total reflected tone. Thus, the power-dependent phase-shift is leveraged to provide gain in a band centered on the circuit's resonance frequency.\\

\begin{figure}[b]
\centerline{\includegraphics[width=12.9cm]{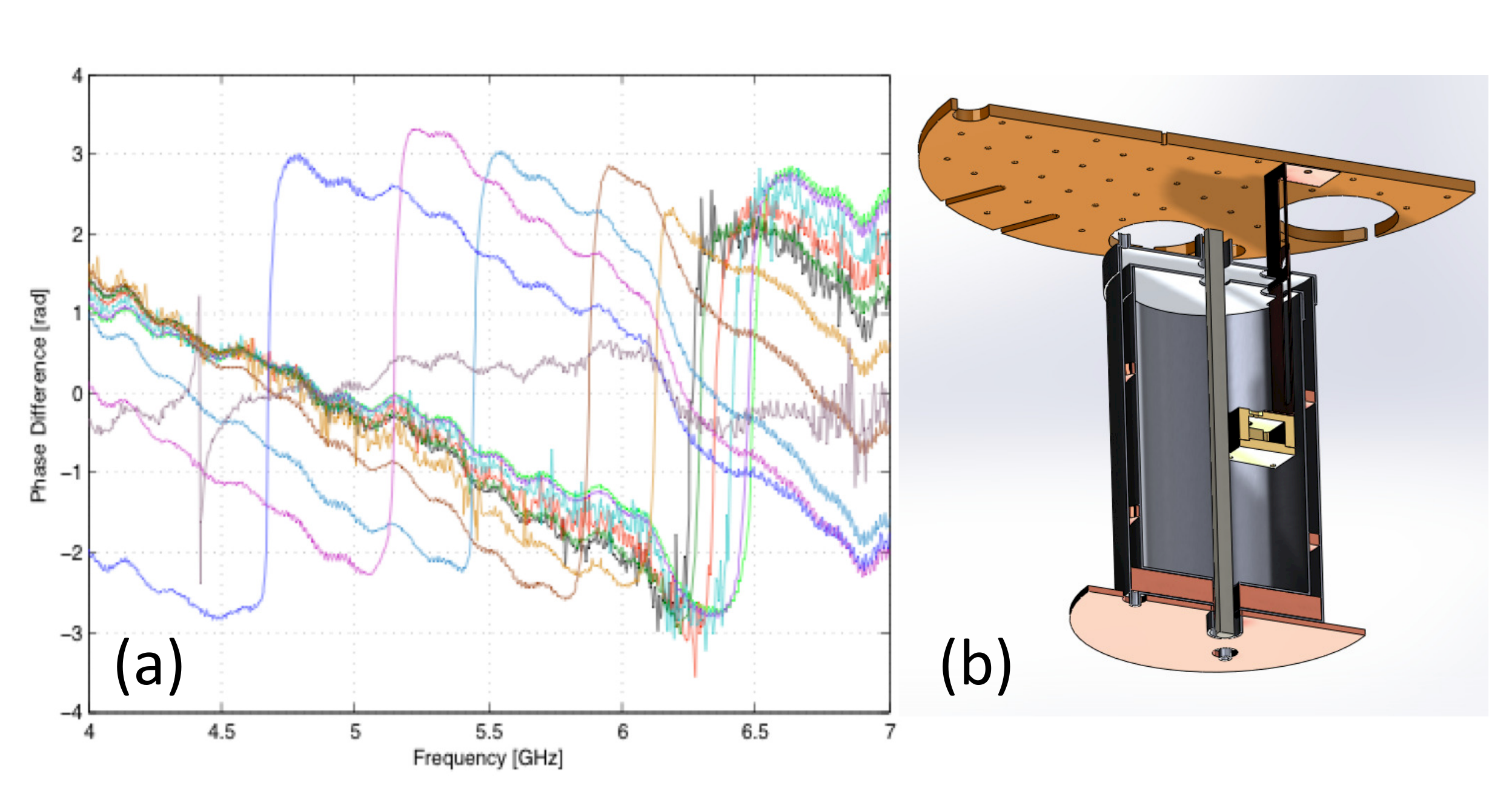}}
%\centerline{\includegraphics[angle =90,width=14.2cm]{IJMPA_Fig_12A.eps}}
\caption{(a) Demonstration of JPA tuning in situ within ADMX-HF.  The center of the $\sim 2\pi$ phase shift of the reflected tone indicates the operating frequency, the range of this device being approximately 4.7 -- 6.5 GHz.  (b) Design of the magnetic shield for the JPA  that resides within the field cancellation coil 40 cm above the top of the main magnet.  }
\label{fig:12}
\end{figure}

\noindent By replacing a single Josephson junction with a pair of Josephson junctions in parallel (i.e. a SQUID), the effective linear part of the Josephson junction inductance increases as magnetic flux threads the resulting SQUID loop. The circuit's resonance frequency can therefore be decreased by introducing flux through the loop (Fig.~\ref{fig:10}).\cite{26} Combining a TKC with two standard microwave components forms a JPA. The resulting amplifiers are narrow band, but the center of the band can be tuned over an octave, 4 -- 8 GHz.  Fig.~\ref{fig:12} shows the magnetic shielding for the JPA as deployed in ADMX-HF and its tuning \emph{in situ}.  Also the JPAs have sufficient gain to realize a quantum-limited microwave measurement.\cite{27}  Furthermore, using the phase-sensitive mode of operation enables a measurement of one quadrature with less noise than the standard quantum limit.\\

\begin{figure}[b]
\centerline{\includegraphics[width=12cm]{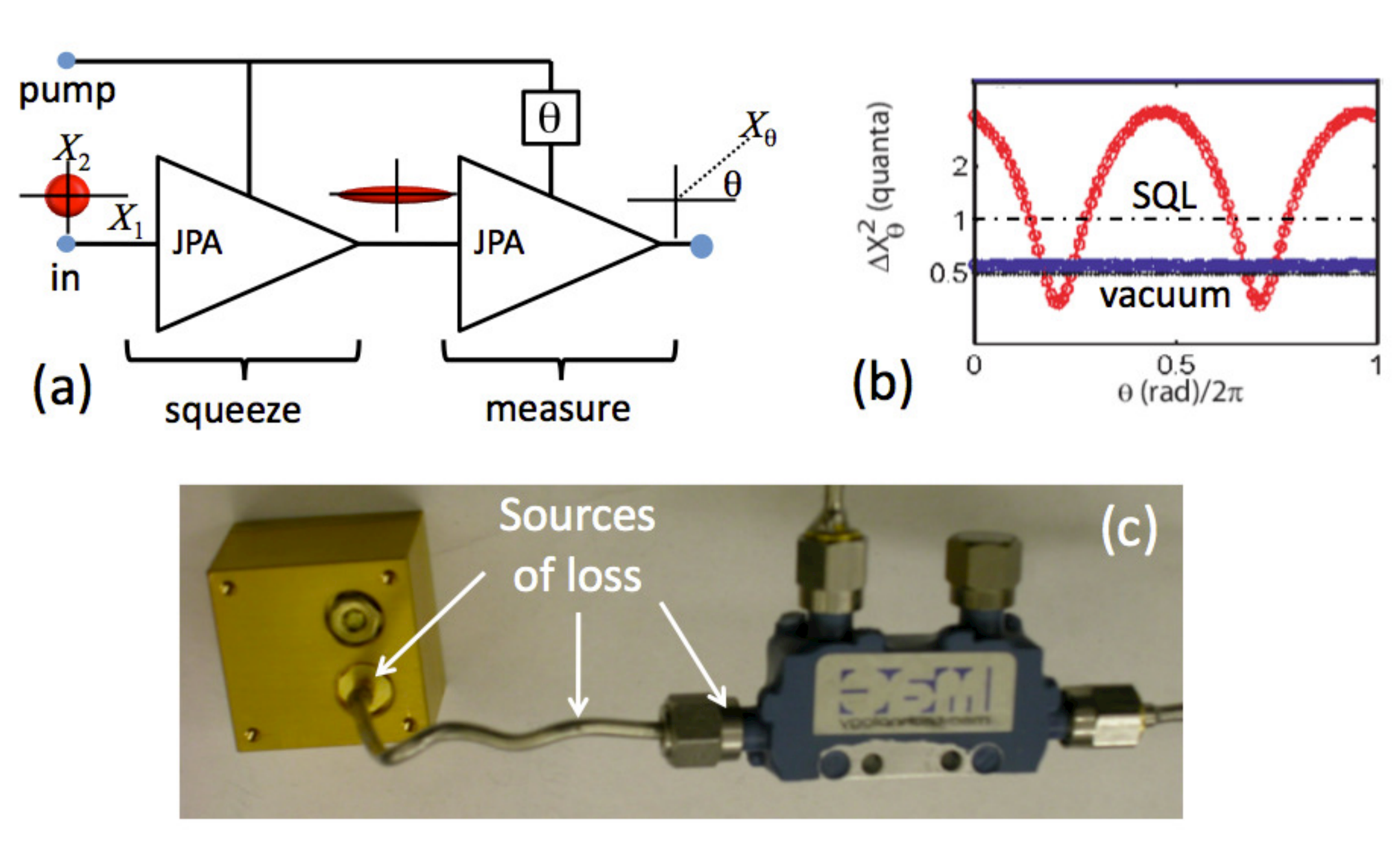}}
%\centerline{\includegraphics[angle =90,width=14.2cm]{IJMPA_Fig_13.eps}}
\caption{Demonstration of noise less than the quantum vacuum.   (a) Microwave vacuum field is fed to a JPA that creates a squeezed state with reduced fluctuations in the sine-quadrature X$_2$ but amplified fluctuations in the cosine-quadrature X$_1$, as illustrated by the phase space cartoons (red circle and ellipse). One quadrature of that squeezed state, chosen by the relative phase between the pumps of the squeezing and measuring JPAs, is measured using a second JPA. (b) The variance of the measured quadrature (red) as its phase is adjusted relative to the squeezed quadrature shows noise not only below the standard quantum limit for measuring vacuum (dashed dotted line) but below the quantum vacuum itself (dotted line). (c) The observed noise is 67$\%$ of vacuum variance; without the loss present in the many cable connections and components it would be less than 10$\%$ of vacuum. The gold box contains a TKC; the blue rectangle is a directional coupler. }
\label{fig:13}
\end{figure}

\noindent The discussion above was limited to JPA performance at the standard quantum limit, which was the original technical baseline for ADMX-HF.  Additionally however, the JILA group has used one JPA to measure the squeezed noise generated by a second JPA (Fig. \ref{fig:13}a,b).\cite{28}  This experiment can be regarded as a proof-of-principle demonstration of a quantum-noise evading axion search.  Here it was not only possible to measure one quadrature below the standard quantum limit, but that quadrature had fluctuations below vacuum because it had been squeezed.\\

\noindent To successfully incorporate squeezed-state techniques into the microwave cavity experiment, it is crucial to minimize losses between the cavity and the amplifier. Although the JILA group's JPA circuits operate as noiseless single quadrature amplifiers, in any of the experiments where signals are transported from one device to a remote JPA amplifier, between 1/2 and 2/3 of the energy is lost.  It is precisely that loss that prevents our measurement of a squeezed state with a noiseless amplifier from exhibiting noise less than 2/3 of vacuum (Fig.~\ref{fig:13}). In transporting the squeezed state (squeezed to 1/20 of vacuum) to the JPA amplifier, 2/3 of the state is absorbed; the vacuum fluctuations associated with that loss replaces the squeezed state. Because the JPA requires a commercial microwave circulator element to separate incident and reflected waves, the JPA cannot be integrated with the signal source, but rather must be fed by external interconnects.  The mitigation strategy will be to reduce this loss to 1/10 by changing all critical signal interconnections from coaxial cables to rigid waveguides. \\

\noindent At frequencies above 16 GHz the basic design of resonant circuit-based and flux-tunable JPAs will likely begin to suffer diminished performance, for two reasons. First, a JJ has a maximum frequency, the plasma frequency, at which it still behaves as a non-linear inductor; this is around 100 GHz for our JJ technology. Using flux to tune the JPA resonance frequency reduces this maximum frequency to between 30 -- 50 GHz. As the operating frequency of the JPA approaches this maximum, the performance of the JPA will suffer. To maintain the necessary gain, bandwidth, and dynamic range of the amplifier, the JPA operating frequency should not exceed 1/3 of the plasma frequency. Second, the transition of a microwave signal from a coaxial cable onto a chip becomes increasingly complex and prone to loss at higher frequencies.  \\

\begin{figure}[h]
\centerline{\includegraphics[width=12cm]{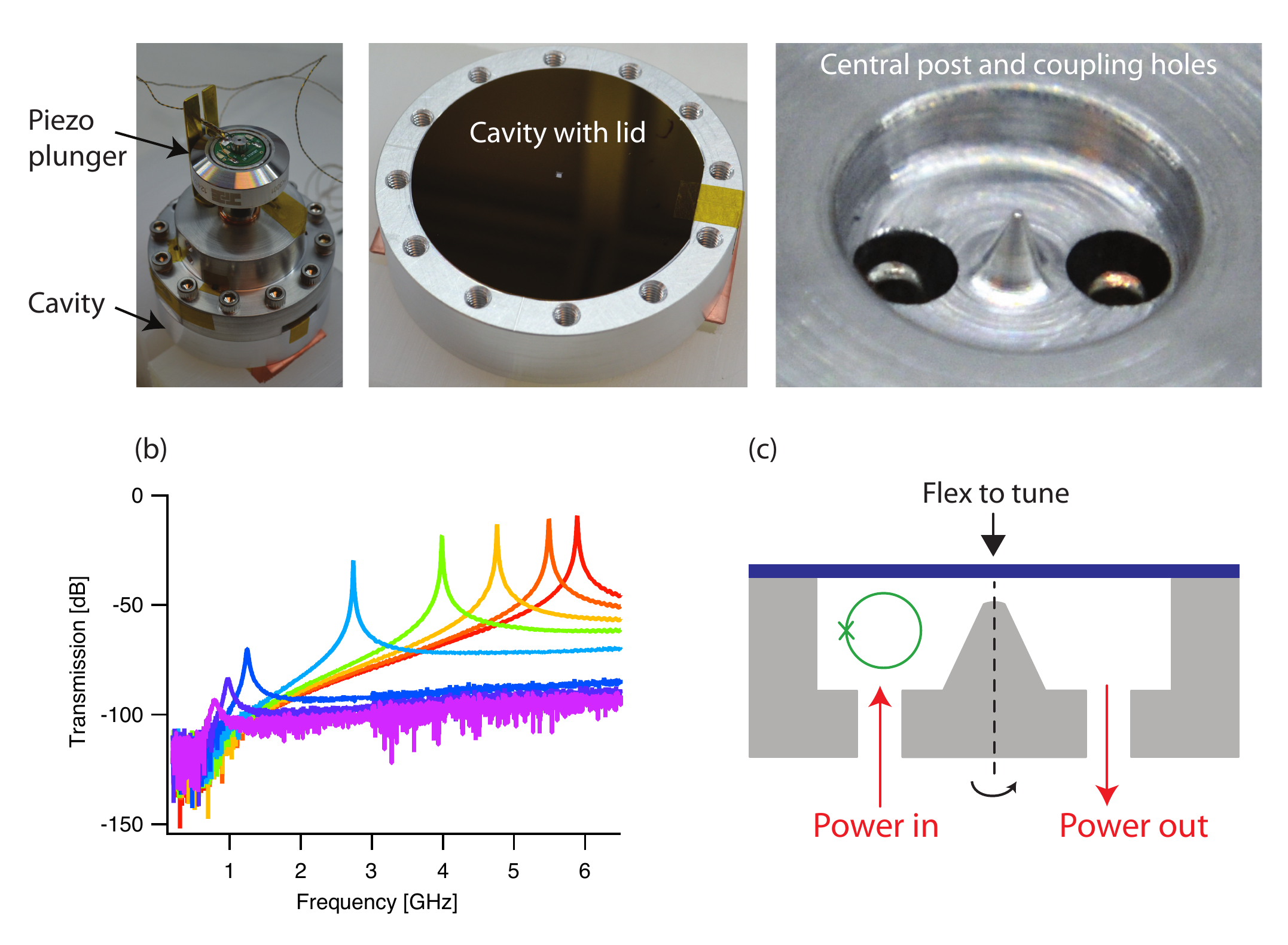}}
%\centerline{\includegraphics[angle=90, width=14.2cm]{IJMPA_Fig_14.eps}}
\caption{ A mechanically tunable cavity. (a) The reentrant cavity from left to right: fully assembled with piezo adjustment, cavity base and lid, cavity base showing the central post and coupling holes.    (b) Transmission through the cavity as a function for frequency for 8 different lid deflections showing the wide mechanical tunability of the cavity. (c) A sketch of a mechanically-tunable reentrant-cavity JPA. The cavity mode magnetic field passes through a superconducting ring (supported in the cavity on a sapphire chip) and experiences a non-linear response due to the JJ interrupting the ring.}
\label{fig:14}
\end{figure}

\noindent The JILA group is developing a new type of JPA for ADMX-HF, consisting of a single JJ contained within a microwave cavity, specifically a reentrant cavity, or Helmholtz resonator. This design overcomes both difficulties: first, by depressing the cylindrical cavity lid towards the central post, the cavity itself is tunable over an octave. As the linear portion of the JJ inductance is no longer used for tuning, the junction will have a 100 GHz plasma frequency. Second, it is much simpler to efficiently couple a cable directly into a cavity than onto a planar circuit. This group has already built superconducting reentrant cavities and tuned their resonance frequencies from 40 GHz down to 8 GHz at 300 mK, without any mechanical feed-throughs into the cryostat. A commercial cryogenically operable stick-slip motor turns a screw, depressing the lid (Fig.~\ref{fig:14}).\\

\noindent In the longer term, we intend to investigate schemes for detecting single microwave photons, which may enable us to circumvent the quantum limit on amplification. If such a technique could be mastered, it could substantially reduce the time required to scan a range of frequency. This is particularly true if the $Q$ of the microwave cavity can be increased, as proposed in the discussion on hybrid cavities above. \\

\noindent In a certain sense, particular quantum circuits called superconducting qubits can detect individual microwave photons.\cite{28} These circuits behave like quantum two-level systems; when they are embedded into microwave cavities, the interaction between the cavity field and the qubit is sufficiently strong that the qubit transition frequency experiences a single-photon Stark-shift that is much greater than any of the dissipative rates. Thus the presence of photons in the cavity can be detected by attempting to promote the qubit to its excited state with a microwave signal in resonance with the zero-photon qubit frequency. If the qubit is not excited, there is at least one photon in the cavity.\\

\noindent While the idea of using qubits to detect the axionic photon flux is appealing, substantial technical barriers must be overcome to realize this concept. Largely because of the high magnetic field, we cannot incorporate the qubit into the axion cavity. Rather the axion cavity must feed a separate cavity that contains the qubit. We must then be able to tune the qubit cavity into resonance with the axion cavity and to read out the state of the qubit at a rate matched to the axion linewidth. Importantly, fast and high-fidelity qubit state readout using JPAs has recently been demonstrated.\cite{29,30} Consequently, in pursuing microwave photodetection we can leverage several technologies (tunable cavities, flexible JPA measurements) already being developed as part of the ADMX-HF project.
\subsection{Summary and conclusions}

Both a strategy and supporting technologies for the microwave cavity search for dark-matter axions  up to 25 GHz ($\sim100\ \mu$eV) appear well in hand.  Some modest development work will be required both for hybrid superconducting cavities and quantum-noise evading devices, such as JPAs operated in squeezed-state mode; much has already been demonstrated on the bench, so the real challenge will be migrating these technologies into the unusual environment of ADMX-HF.  \\

\section*{Acknowledgments}

This work was supported under the auspices of the National Science Foundation, under grants PHY-1067242, and PHY-1306729, and the auspices of the U.S. Department of Energy by Lawrence Livermore National Security, LLC, Lawrence Livermore National Laboratory under Contract DE-AC52-07NA27344.

%%%%%----------------------------------	

%\section*{References}

%References are to be listed in the order cited in the text in Arabic
%numerals.  They should be listed according to the style shown in the
%References. Typeset references in 9 pt roman.

%References in the text can be typed in superscripts,
%e.g.: ``$\ldots$ have proven\cite{autbk,edbk,rvo} that
%this equation $\ldots$'' or after punctuation marks:
%``$\ldots$ in the statement.\cite{rvo}'' This is
%done using LaTeX command: ``\verb|\cite{name}|''.

%When the reference forms part of the sentence, it should not
%be typed in superscripts, e.g.: ``One can show from
%Ref.~\refcite{autbk} that $\ldots$'', ``See
%Refs.~\citen{jpap,colla,autbk}, \refcite{rvo}
%and \refcite{pro} for more details.''
%This is done using the LaTeX
%command: ``Ref.~\verb|\citen{name}|''.

%\begin{thebibliography}{000} %for 3 digits

\end{document}